\documentclass{aa}  
\usepackage{graphicx}
\usepackage{txfonts}
\usepackage{longtable}
\begin{document}
\newcommand{\water}{H$_2$O}
\newcommand{\micron}{$\mu$m}
\newcommand{\kms}{km\,s$^{-1}$}
\newcommand{\mdot}{$\dot{\rm{M}}$}
\newcommand{\Myr}{$M_{\sun}$\,yr$^{-1}$}
   \title{Lifetime of OH masers at the tip of the asymptotic giant branch}


   \author{D. Engels
          \inst{1}
          \and
          F. Jim\'enez-Esteban\inst{1}\fnmsep\thanks{Present address:
Observatorio Astronomico Nacional, Apartado 112,
           E--28803 Alcala de Henares, Spain}
          }

   \offprints{D. Engels}

   \institute{Hamburger Sternwarte, Gojenbergsweg 112,
           D--21029 Hamburg, Germany\\
              \email{dengels@hs.uni-hamburg.de}
             }

   \date{Received ???; accepted ???}

 
  \abstract
     {A large fraction of otherwise similar asymptotic giant branch stars 
     (AGB) do not show 
     OH maser emission. As shown recently, a restricted lifetime
     may give a natural explanation as to why only part of any sample emits
     maser emission at a given epoch.} 
   {We wish to probe the lifetime of 1612 MHz OH masers in circumstellar shells
of AGB stars.}
   {We reobserved a sample of OH/IR stars discovered more than 28 years ago
    to determine the number of stars that may have since lost their masers.}
   {We redetected all 114 OH masers. The minimum lifetime inferred
is 2800 years (1$\sigma$). This maser lifetime applies to AGB stars
with strong mass loss leading to very red infrared colors. The velocities and
mean flux density levels have not changed since their discovery. As the
minimum lifetime is of the same order as the wind crossing time,
strong variations in the mass-loss process affecting the excitation 
conditions on timescales of $\approx$3000 years or less are unlikely.} 
   {}

   \titlerunning{Lifetime of OH masers at the tip of the AGB}
   \maketitle
%

\section{Introduction}
Molecular OH maser emission at 1612 MHz is frequently observed from
stars approaching the tip of their evolutionary track on the
asymptotic giant branch (AGB). These stars are in general long-period
variables with pulsation periods $>$1~year.  The pulsation is
accompanied by heavy mass loss, which forms a circumstellar envelope
(CSE) of gas and dust, and if the mass-loss rate surpasses \mdot\
$\approx$10$^{-6}$ \Myr\ the dust envelope becomes opaque for visible
light. Classically these `invisible' stars were called OH/IR stars,
but nowadays this term is often used in general reference to OH
emitting AGB stars selected in the infrared (see review by Habing
\cite{Habing96}).

Interestingly, a large fraction ($\approx$40\%) of any sample of IRAS
sources, selected to match the infrared colors of known OH/IR
stars, does not exhibit a detectable OH 1612 MHz maser (Lewis
\cite{Lewis92}).  These infrared sources were baptized by Lewis as `OH/IR
star color mimics'.  Some of them do exhibit mainline OH (at 1665 or
1667 MHz) and/or 22 GHz \water\ masers (Lewis \& Engels
\cite{Lewis95}), corroborating their cousinhood with OH/IR stars.
OH/IR stars are therefore only part of the population of evolved and
obscured AGB stars with oxygen-rich chemistry.

The reasons for the absence of 1612 MHz OH masers in a large fraction
of oxygen-rich AGB stars are not known.  The maser photons are emitted
by a transition between hyperfine levels of the groundstate of OH,
which are inverted with the help of pump photons at $\lambda$=35 and
53\micron, emitted from dust in the CSEs. A requirement for
excitation is sufficient OH column density, which might be low in
mimics due to the destruction of OH by the interstellar UV field. This
effect cannot account for mimics in general, as mimics are also
present at higher galactic latitudes, where the influence of the UV
field is low. In addition, the presence of a hot white dwarf companion
leading to photodissociation of OH is not generally able to suppress
the OH maser (Howe \& Rawlings \cite{Howe94}). A further requirement
is velocity coherence over large distances to allow
amplification. Turbulence may disrupt this coherence in the case of
mimics.

An alternative explanation is the assumption that the OH maser is only
present temporarily on the AGB and therefore these stars may change
between an OH/IR status and that of a mimic. This explanation has been
triggered by recent observations showing the fading of the OH maser in
\object{IRAS~18455+0448} over a time range of a decade (Lewis et
al. \cite{Lewis01}) and the absence of masers in four stars during a
revisit of 328 OH/IR stars 12 years after their first detection (Lewis
\cite{Lewis02}).  The high rate of `dead' OH/IR stars among a
sub-sample of 112 OH/IR stars with relatively blue colors led Lewis to
conclude that the mean 1612 MHz emission life is in the range 100--400
years.

To test this lifetime we reobserved another sample of N$>$ 100 OH/IR stars,
which was drawn from the first surveys for OH maser emission prior
to 1980. With a difference of almost 30 years between the two observations
several masers were expected to have disappeared.

\section{Observations}
\addtocounter{table}{1} The sample of OH/IR stars was mainly taken 
from the compilation of Baud et al. (\cite{Baud81a}), which contains
N=114 OH maser sources located at $10^{\circ} \le l \le 150^{\circ},
\mid b\mid \le 4.2^{\circ}$. Their source, \object{OH~57.5+1.8}, was
found by Engels (\cite{Engels96b}) to be a blend of two OH masers and
is included here as two objects, OH 57.5+1.8A and B. In addition,
\object{OH~19.2-1.0} was added, which was omitted by Baud et
al. because of its peculiar maser profile. The full sample contained
N=116 objects and is listed in Table \ref{sample}. The compilation of
Baud et al.  contains only masers that were discovered before 1978.

To ensure that a non-detection would not be caused by inaccurate
coordinates, we compiled the literature for the best available radio
coordinates and coordinates for their infrared counterparts.  The
original coordinates could have been wrong by several arc minutes, and
not for all sources improved radio coordinates were later obtained by
follow-up observations. Infrared counterparts (IRAS, MSX) were found
for N=113 OH maser sources, while there are no convincing counterparts
for \object{OH~18.3+0.1}, \object{OH~39.6+0.9}, and
\object{OH~42.8$-$1.0}. In the case of OH~18.3+0.1, the absence of an
infrared counterpart coincident with the accurate position given by
Bowers \& de Jong (\cite{Bowers83a}) confirms the suspicion of
Winnberg et al. (\cite{Winnberg81}) that this maser is actually of
interstellar nature. OH~39.6+0.9 was discovered by Johansson et
al. (\cite{Johansson77}), which gave an error of rms=$2\farcm5$ for
its position. The source was never reobserved. The nearest IRAS
counterparts are \object{IRAS~18578+0616} and \object{18584+0616},
with a distance of $4\farcm3$. \object{OH~42.8$-$1.0} has also never
been reobserved since its discovery by Baud et
al. (\cite{Baud79b}). The positional error of the discovery position
is rms$\approx$3\arcmin. \object{IRAS~19108+0815} is the nearest IRAS
counterpart, with a distance of $2\farcm4$. The IRAS source is weak
and was not detected by MSX. We also suspected that OH~39.6+0.9 and
OH~42.8$-$1.0 are not OH/IR stars. Nevertheless, all three sources
were reobserved.


The observations were made during two nights 2005, August 2-4 with
the Effelsberg radio telescope.  As frontend we used the 1.3-1.7 GHz
HEMT receiver and as backend an 8192 channel autocorrelator. The
correlator was split into four segments, of which three were centered
on the two OH main line frequencies 1665.401 and 1667.359 MHz and one
on the OH satellite line frequency 1612.231 MHz. The fourth segment
was not used because of technical problems.  In this paper we focus on
the results of the 1612 MHz observations, for which the receiver
passed left circular polarization. We chose a bandwidth of 1.25~MHz,
yielding a velocity resolution of 0.11 \kms\ and a velocity coverage
of $\pm$112.5 \kms, centered on the mean velocity of the two OH maser
lines, as given by Baud et al. (\cite{Baud81a}).  The beamwidth was
$7\farcm8$. The coordinates to which the telescope was pointed are
given in Table \ref{sample}. If not otherwise stated, they were taken
from the MSX6C catalog. The errors of all coordinates are smaller than
a few arcsec, and therefore much smaller than the beamwidth.  System
temperatures were $\approx$25~K at zenith. We observed in position
switch mode
with an integration time of 6 minutes (ON+OFF), yielding a typical
sensitivity of 0.25 Jy (3$\sigma$).

   \begin{figure}[ht]
   \centering
   \includegraphics[width=8.5cm]{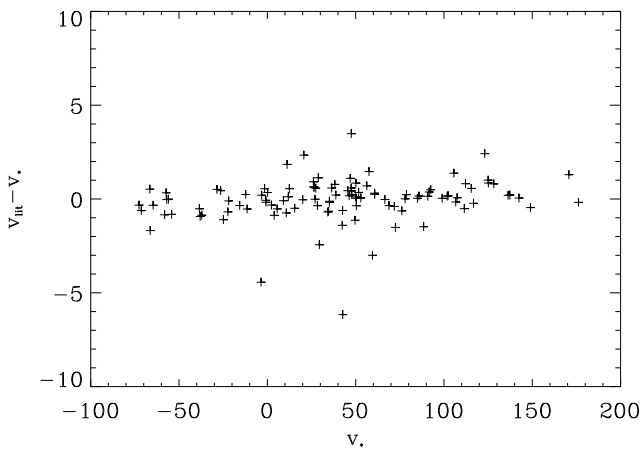}
      \caption{Deviations of measured radial velocities v$_\star$ 
               of the OH/IR stars observed from radial velocities v$_{lit}$
               given by Baud et al. (\cite{Baud81a}).
              }
         \label{velocities}
   \end{figure}

The data reduction was made within CLASS and included removal of the
baseline and flux calibration. The calibration was made against 3C286
and 3C48 adopting flux densities of $13.53\pm0.05$ Jy and
$14.25\pm0.05$ Jy, respectively (Ott et al. \cite{Ott94}). No gain
curve was applied as the flux calibrators were observed several times
during the nights and showed no elevation dependent variations over the
$20 - 70^\circ$ elevation range observed.

   \begin{figure}[ht]
   \centering
   \includegraphics[width=5cm]{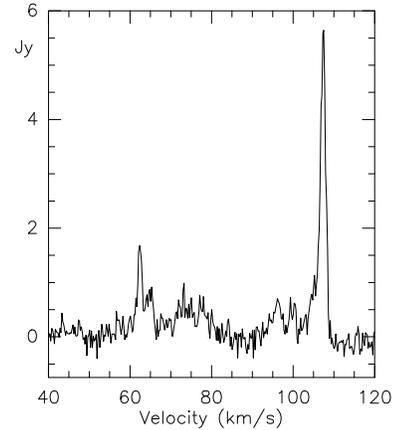}
      \caption{1612 MHz OH maser spectrum of \object{OH~24.7$-$0.1}.
              }
         \label{OH24}
   \end{figure}

\section{Results}
We recovered the masers for all 113 OH maser sources with infrared
counterparts.  In general, velocities and flux densities were
compatible with the values given by Baud et
al. (\cite{Baud81a}). Table \ref{sample} lists the ends v$_b$ and
v$_r$ of the full velocity range, over which emission was detected,
and the integrated flux densities SI$_b$ and SI$_r$. We adopted the
midpoint of the velocity range v$_b -$v$_r$ as radial velocity
v$_\star$ of the star. The integrated flux densities SI$_b$ and SI$_r$
were determined by integrating the calibrated spectra over the two
halves of the velocity range v$_\star -$v$_b$ and v$_r -$v$_\star$.

Figure \ref{velocities} shows a comparison of the velocities between
our data and the values listed by Baud et al. (\cite{Baud81a}).
Within the errors ($\pm2$ \kms) there is good agreement between the
velocities. In a few sources, weak emission was found outside the main
peaks, which led to radial velocities deviating by as much as 7 \kms\
from the previous values. \object{OH~24.7$-$0.1} and \object{OH~20.4$-$0.3} 
are not included in the figure. In the case of
\object{OH~24.7$-$0.1} (cf. Fig. \ref{OH24}), the velocities needed a
complete revision. In the discovery paper of this source (Johansson
et al. \cite{Johansson77}), only the stronger of the two lines at
$\approx$111 \kms\ was clearly detected, while a marginal feature at
$\approx$142 was considered as the second line. The new spectrum shows
that the second line is at the lower velocity of 62 \kms. In
\object{OH~20.4$-$0.3} we could not measure v$_*$ because the blue
emission peak was corrupted by interference.

   \begin{figure}[ht]
   \centering
   \includegraphics[width=8.5cm]{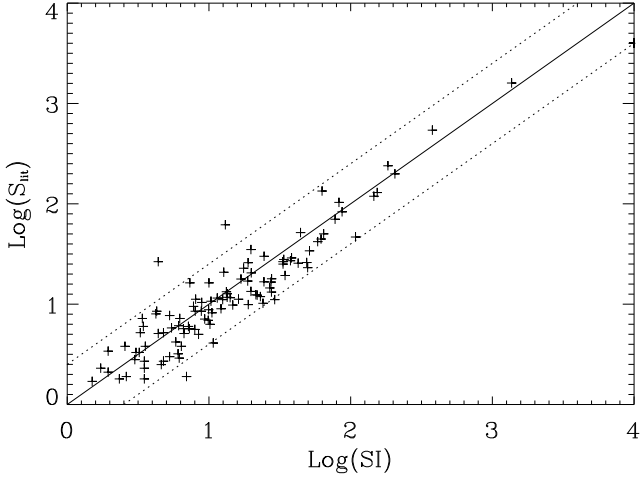}
      \caption{Comparison of measured integrated flux densities 
               SI =  SI$_b +$SI$_r$ 
               of the OH/IR stars observed with flux densities S$_{\rm{lit}}$ 
               given by Baud et al. (\cite{Baud81a}). The units
               are $10^{-22}$ Watt m$^{-2}$. The dashed lines delimit
               the deviations expected due to flux variations by a factor
               2.5. 
              }
         \label{fluxes}
   \end{figure}

Figure \ref{fluxes} shows a comparison of the integrated flux
densities.  The scatter is largely within the margins expected from
the large amplitude variations due to the pulsations of the
star. Typical variations have a factor of about 2.5 (Herman \& Habing
\cite{Herman85}).

\begin{figure}[ht]
   \centering
   \includegraphics[width=5cm]{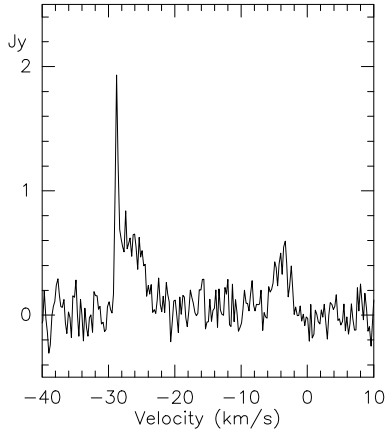}
      \caption{1612 MHz OH maser spectrum of \object{OH~42.8$-$1.0}.
              }
         \label{OH42}
   \end{figure}

Mixed results were obtained for the three maser sources without
IRAS/MSX counterparts. For \object{OH~18.3+0.1} we found a single line
at 9.8 \kms\ (3 Jy), which possibly belongs to the interstellar maser
line at +11 \kms, discussed by Baud et al. (\cite{Baud79b}). For
\object{OH~39.6+0.9} we searched for the maser in a 15\arcmin
x15\arcmin\ region around the position of the two nearest IRAS
sources, after no OH masers were detected at the IRAS positions
themselves. No maser could be found down to 0.8~Jy
(3$\sigma$). \object{OH~42.8$-$1.0} was detected (Fig.
\ref{OH42}). To improve the coordinates we tried to maximize the
signal. The improved position is $\alpha$ = 19h 13m 36.4s ($\pm$5.0s),
$\delta$ = +08\degr\ 22\arcmin\ 39\arcsec\ ($\pm$2\arcmin).  The
position is still too coarse to allow an unambiguous identification
with an infrared source. The most likely candidate is
\object{IRAS~19112+0816}, which is approximately 1\arcmin\ away.

 To summarize, the only sources from the sample of OH/IR stars of Baud
 et al. (\cite{Baud81a}), that we did not redetect at all, or not at
 the expected velocities, are \object{OH~18.3+0.1} and
 \object{OH~39.6+0.9}. They are two of the three objects for which no
 infrared counterpart was found.  We therefore conclude that these two
 sources are not OH/IR stars, while for the third,
 \object{OH~42.8$-$1.0}, the classification as OH/IR star is
 retained. We further summarize that all OH/IR stars in the sample (N=113
 + OH~42.8$-$1.0) still possess an 1612 MHz OH maser. As these stars
 were discovered prior to 1978, we adopt $\Delta t = 28$ years as the
 minimum time passed between discovery and redetection in 2005.

\section{Lifetime of OH Masers}
The restrictions on the lifetime of OH masers following from the
redetection of all bona-fide OH/IR stars, can be accessed using a
basic law of combinatorics.

We will assume that stars enter at random times into the phase where
they support maser emission. We assume also that the mean lifetime of
an OH maser in a CSE is $T$ years, and a known OH maser is revisited
after $\Delta t$ years and has disappeared. Assuming $\Delta t < T$,
the probability 
to disappear is $\Delta t / T$ and the probability to detect a
maser again 
is $(1-\Delta t / T)$.

If a sample of $n$ OH masers is revisited and all are redetected, the
accumulated probability $P^0_n$ that no maser disappeared within
$\Delta t$ years is
\begin{equation}
P^0_n = \left(1 - \frac{\Delta t}{T} \right)^n
\end{equation}

The probability to detect the disappearance of a single maser is
\begin{equation} \label{one-eq}
P_n^1 = n \cdot \left(\frac{\Delta t}{T} \right) \cdot  \left(1 - \frac{\Delta t}{T} \right)^{n-1}
\end{equation}
where the factor $n$ allows for the fact that any of the $n$ masers might have
been affected.

More general, the probability to detect the disappearance of $m$ masers among
$n$ stars after $\Delta t$ years is
\begin{equation} \label{combi-eq}
P_n^m = \frac{n!}{m! (n-m)!} \cdot \left(\frac{\Delta t}{T} \right)^m \cdot  \left(1 - \frac{\Delta t}{T} \right)^{n-m}
\end{equation}
This general equation (a standard relation in combinatorics) includes
the case that no maser ($m=0$) disappeared, or that all of them are
extinguished ($m=n$), e.g.,  the valid range for $m$ is $0\le m\le n$.

   \begin{figure}[ht]
   \centering
   \includegraphics[width=6cm, angle=-90]{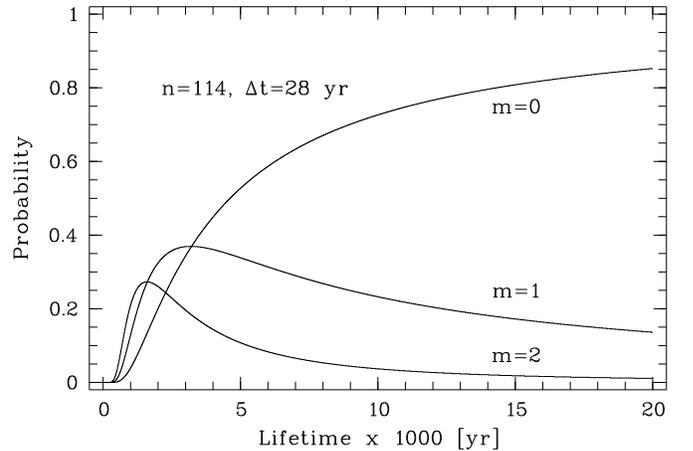}
      \caption{Probabilities $P_n^m$ that out of $n=114$ OH masers, $m$
               masers with  a lifetime $T$ will
               have disappeared after $\Delta t = 28$ years.
              }
         \label{lifetime0-2}
   \end{figure}

   In Fig. \ref{lifetime0-2} we plot the probabilities derived from
   Eq. (\ref{combi-eq}) for the case $n=114$ stars, $m=0-2$ missing
   masers, and $\Delta t=28$ years for lifetimes $T$ up to 20\,000
   years. The curves tell us for $T=15\,000$ years, for example, that the
   probability to redetect all masers is 81\%, and that there is a 17\%
   chance of observing one star with an extinguished maser. Even
   the non-detection of two masers is non-negligible: 2\%. For smaller
   lifetimes, say $T<5000$ years, the 
   cumulative probability for $m \le 2$ does not approach
   100\%, because now the disappearance of $m>2$ masers is
   not unlikely anymore. 
   
   \begin{table}
      \caption{Lower lifetime limits  
               $T_{\rm{min}}$ at different significance levels
               for 1612 MHz OH masers
               based on observations in this paper. $P_n^m$ is the 
               probability for finding exactly $m$ extinguished maser among
               $n$ stars reobserved after $\Delta t$ years.}
         \label{lA-prob}
	 \centering
         \begin{tabular}{rrrrrll}
            \hline\hline
            \noalign{\smallskip}
             $n$ & $m$ &$\Delta t$& $T_{\rm{min}}$& $P_n^m$ & \multicolumn{2}{l}{Comments} \\
            \noalign{\smallskip}
                 &     & [yr]     & [yr]    &  [\%]   &      &        \\
            \noalign{\smallskip}
            \hline
            \noalign{\smallskip}
            114  &  0  &  28      &    550  &  0.2  &  3$\sigma$ & This paper\\
            114  &  0  &  28      &   1050  &  4.6  &  2$\sigma$ &\\
            114  &  0  &  28      &   2800  & 31.8  &  1$\sigma$ &\\
            \noalign{\smallskip}
            \hline
         \end{tabular}
   \end{table}

   As expected, the probability to redetect all masers decreases
   towards lower lifetimes. 
   There is no absolute lower lifetime limit, but it is possible
    to define such limits in relation to significance levels. 
   For example, lifetimes $T$ in which the probability of finding 
   {\it at least} one maser extinguished exceeds 68.2\%, can be
   excluded at the 1$\sigma$ level. The probability of finding 
   {\it at least} one extinguished maser is ($1 - P^0_{114}$), which
    is different from the probability $P_{114}^1$ (Eq.
   \ref{one-eq}) for finding {\it exactly} one extinguished maser. 
   For our sample, probabilities $(1 - P^0_{114}) >
   0.682$  
   or $P^0_{114} < 0.318$ to detect an extinguished maser are obtained  
   for lifetimes $T \la 2800$ years (Fig. \ref{lifetime0-2}). 
   As all masers were redetected, $T_{\rm{min}} = 2800$ years is a lower 
   limit of the lifetime at the 1$\sigma$
   level. Lower lifetime limits with higher significance levels are given in
   Table \ref{lA-prob}. Lifetimes of T$<$400 years, as derived by Lewis
   (\cite{Lewis02}), have P$^0_{114}<0.01$ and can be excluded for our
   sample.

\section{Discussion}
The high rate of extinguished masers in the sample studied by Lewis
(\cite{Lewis02}) (Arecibo sample) is at odds with our results. The
rate is also astonishing given that previous monitoring programs did
not witness similar maser luminosity fading processes over periods of
several years. On the other hand, a disappearance of masers in our
sample would not have been too surprising. For one part of the sample, 
drastic mass-loss variations due to the onset of a `superwind' a few
hundred years ago was postulated (Justtanont et
al. \cite{Justtanont06}), and another part of the sample belongs to the
group of `non-variable OH/IR stars', for which a loss of their masers
within a couple of thousand years is predicted (Engels
\cite{Engels02}).


\subsection{Lifetime limit derived from  the Arecibo sample}
Before discussing possible causes due to the different sample
properties, we will have a look at the restrictions on lifetimes for
the Arecibo sample using the equations of the previous section.  Lewis
(\cite{Lewis02}) concluded that after $\Delta t=14$ years, from 
$n=112$ stars, five masers will have disappeared. 
The affected stars are the IRAS objects 15060+0947, 18455+0448,
19479+2111, 19529+3634, and 20547+0247.
  He argues that the
most likely lifetime is $T=314 (-97,+387)$ years. The maser in
\object{IRAS~15060+0947}, which was steadily declining at the time of
the Lewis (\cite{Lewis02}) publication, actually never completely disappeared. 
Another one, \object{IRAS~19479+2111}, reappeared in 2005
(Lewis, priv.  communication). The probable lifetime of masers in the
Arecibo sample is therefore certainly longer than the T=314 years.


We will ask here the probability of finding $m$ or more extinguished
masers for a given lifetime. This allows the determination of upper limits for
the OH maser lifetimes in the Arecibo sample at different significance
levels. For the purpose of this calculation, we will use $m=4$
extinguished masers (omitting \object{IRAS~15060+0947}) and make an
adjustment to $\Delta t$=12 years.

The probability of finding at least $m$ masers extinguished among
$n$ stars after $\Delta t$ years is
\begin{equation} \label{cumul-eq}
Q_n^m = \sum_{i=m}^n P_n^i
\end{equation}
For $m=1$, the case discussed in the previous section, one finds
\begin{equation}
Q_n^1 = \sum_{i=1}^n P_n^i = \sum_{i=0}^n P_n^i - P_n^0 = 1 -  P_n^0
\end{equation}

\begin{figure}[ht]
   \centering
   \includegraphics[width=6cm, angle=-90]{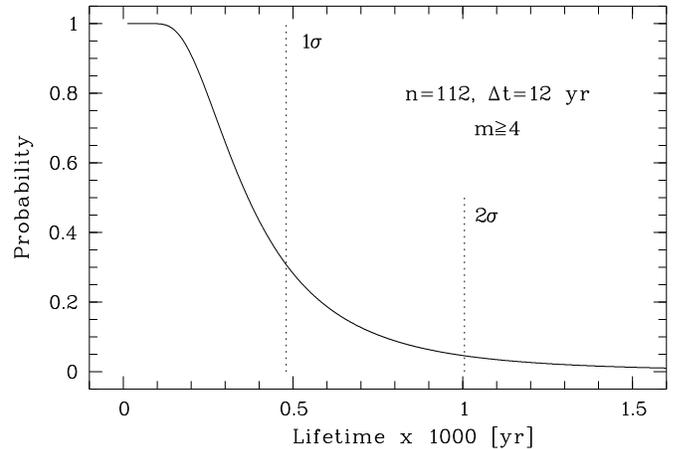}
      \caption{Probabilities $Q_{112}^4$ that out of $n=112$ OH
               masers, m$\ge$4 masers with a lifetime $T$ will have
               disappeared after $\Delta t = 12$ years.  The vertical
               lines mark the upper lifetime limits $T_{\rm{max}}$ at
               the 1$\sigma$ and 2$\sigma$-level significance.  }
         \label{lifetime5plus}
   \end{figure}

The probability $Q_{112}^4$ is plotted as a function of $T$ in
Fig. \ref{lifetime5plus}. This probability drops below 31.2\% at
$T=476$ years, meaning the exclusion of longer lifetimes at the
1$\sigma$-level.  Lifetimes $T\ga2400$ years can be safely excluded, as
the probability falls below 0.2\% (3$\sigma$) (Table \ref{lB-prob}).

   Thus, the high rate of extinguished masers in the Arecibo sample
   indeed points to rather short lifetimes, while in our sample,
   similar short lifetimes are unlikely. However, as Lewis
   (\cite{Lewis02}) already pointed out, the two samples cannot be
   compared directly, because the stars with extinguished masers
   mostly belong to a population of OH/IR stars with low main-sequence
   masses, blue IRAS colors, periods $P<700$ days, and envelope
   expansion velocities $v_e < 12$ \kms, which have to be
   distinguished from the {\it classical} OH/IR stars observed here.
   The latter have larger progenitor masses, redder IRAS colors,
   periods 1000--2000 days, and $v_e > 12$ \kms.  The classical OH/IR
   stars have larger mass-loss rates and probably create CSEs,
   providing an environment more favorable for OH maser emission. The
   higher stability of their masers may then be responsible for the
   longer lifetimes compared to those hosted by the less dense
   envelopes of the bluer OH/IR stars.

   \begin{table}
      \caption{Upper lifetime limits 
             $T_{\rm{max}}$ at different significance levels
               for 1612 MHz OH masers, based on 
               observations of the Arecibo sample. $Q_n^m$ is the 
               probability for finding at least $m$ extinguished masers among
               $n$ stars reobserved after $\Delta t$ years.}
         \label{lB-prob}
	 \centering
         \begin{tabular}{rrrrrll}
            \hline\hline
            \noalign{\smallskip}
             $n$ & $m$ &$\Delta t$& $T_{\rm{max}}$ & $Q_n^m$ & \multicolumn{2}{l}{Comments} \\
            \noalign{\smallskip}
                 &     & [yr]     & [yr]    &  [\%]    &     &  \\
            \noalign{\smallskip}
            \hline
            \noalign{\smallskip}
            112  &  4  &  12 &  2410 &  0.2  &  3$\sigma$ & Arecibo   \\
            112  &  4  &  12 &  1005 &  4.6  &  2$\sigma$ & sample \\
            112  &  4  &  12 &   476 & 31.2  &  1$\sigma$ &   \\
            \noalign{\smallskip}
            \hline
         \end{tabular}
   \end{table}

\subsection{Monitoring programs}
In line with our results, thus far, no extinction of an OH maser has
been reported from monitoring programs. The longest was carried out
over 15 years by Etoka \& Le Squeren (\cite{Etoka00}), but contained
only 7 Mira variables, which are IRAS sources with blue colors.  A
larger number of OH/IR stars (n=37) was covered by van Langevelde et
al. (\cite{Langevelde93}), however the monitoring time was only 3
years. 1$\sigma$ lower limits $T_{\rm{min}} \approx100$ years for
these samples do not give useful lifetime constraints.  A larger
sample containing 60 OH/IR stars was monitored over 10 years by Herman
\& Habing (\cite{Herman85}) and van Langevelde et al.
(\cite{Langevelde90}), but it is a subsample of the present one
studied here, and therefore gives no independent information.

\subsection{A recent onset of a `superwind'?}
Drastic changes of the excitation conditions for OH masers on
timescales of hundreds of years is implied by the two wind regime, 
invoked by Justtanont et al. (\cite{Justtanont96}) to model the CSE
of the classical OH/IR star \object{OH 26.5+0.6}. This object is a
prototype OH/IR star and part of our sample. Their model consisted of
an inner `superwind' and an outer more tenuous AGB wind (hosting the
OH maser).  The transition to the 550 times higher mass-loss rates of
the `superwind' lasted less than 150 years and started $\approx$200
years ago. Support for a generalization of this model to OH/IR stars
has commonly been obtained from observations of water-ice
(Justtanont et al. \cite{Justtanont06}).  Water-ice is formed in the
outer envelopes at radii $>10^{16}$ cm, and the difference between
OH/IR stars with or without a water-ice absorption band at 3.1\micron\
might be related to the question, if the high dust densities of the
`superwind' have already reached this region or not.  Given the short
timescales involved, one expects that, at least in one star, the
transition region between both winds has passed its OH maser shell
during the last 30 years. However, no evidence comes from the flux
level of the OH masers or their velocities that the excitation
conditions of any of them (encompassing most of the stars studied by
Justtanont et al.)  have been changed due to a several hundred-fold
increase of the density or the number of pump photons in their
surroundings.

\subsection{Non-variable OH/IR stars}
For part of our sample, a lifetime of the OH masers of less than a few
thousand years is predicted due to the dissipation of the CSE during
transition to the post-AGB phase. Stars in this phase are
distinguished by their non-variability and are called `non-variable
OH/IR stars', with non-variability defined as absence of
large-amplitude variations typical for Mira stars. The non-variable
OH/IR stars are thought to have stopped their pulsations and the
associated strong mass loss, and have therefore developed a hollow
shell, which continues to expand.  Inside, the densities will
have dropped sharply and it takes $\la$2000 years for the inner
boundary of this shell to reach the OH maser region, assuming
a mean expansion velocity 12 \kms\ and a OH maser shell radius of
several $10^{16}$ cm. The drop in density will 
ultimately
 extinguish the
OH masers. Support for this scenario was given by Engels
(\cite{Engels02}), who reported the disappearance of the \water\ maser
of \object{OH~17.7-2.0} (a present sample member) within a
decade. This extinction of the maser was attributed to the expected
drop of densities in the \water\ maser zone, after which the inner
boundary of the hollow shell passed. As \water\ masers in OH/IR stars
are typically located $\approx$10 times closer to the star than OH
masers, the extinction of the OH maser in OH~17.7-2.0 is expected to
happen in the coming centuries. Also, the disappearance of the OH
maser in \object{IRAS~18455+0448}, an Arecibo source with very red
infrared colors, can be explained by its possible status as an
emerging post-AGB star (Lewis et al. \cite{Lewis01}).

Among our sample, 45 OH masers were monitored by Herman \& Habing
(\cite{Herman85}) and they found 13 ($\approx$30\%) of them with no or
very weak (irregular) variations. Assuming that this fraction of
non-variable objects is valid for our full sample, we find that the
lifetime of the OH masers in non-variable OH/IR stars is $>$820 years
(1$\sigma$). The re-detection of all OH masers is therefore still
compatible with the classification of these stars as transition
objects.  On the other hand, Gray et al. (\cite{Gray05}) modeled the
decline of the OH maser emission after detachment of the CSE and found
that the masers disappear even before the inner border of the detached
envelope has reached the OH maser shell, because they depend on
pumping photons that emerge from the dust further inside. Therefore,
the decay of the masers starts immediately after detachment and is
finished in their models within $<$100 years. To maintain the
classification of the non-variable OH/IR stars as post-AGB stars, it
is therefore mandatory to assume that the mass loss does not stop
abruptly with the cessation of the large-amplitude variations at the
end of the AGB evolution. The mass-loss rates more likely decline
gradually over a time range of several thousand years.

\subsection{Envelope properties and the 
stability of OH masers}
Because of the rarity of masers in the post-AGB phase, masers in
general will disappear with the end of AGB evolution. The OH/IR stars
with bluer envelopes are however genuine AGB stars and therefore, the
disappearance of their masers requires a different explanation.  The
traditional assumption is that the timescales for which mass-loss rates
change in response to evolutionary changes in the AGB (the
intermittent thermal pulses) are much longer than the wind crossing
times of a few thousand years through the CSE. Prior to departure from
the AGB, there are only the phases during and after a thermal pulse in
which shorter timescales are prevalent.  Lewis (\cite{Lewis02})
therefore associates the observed short lifetimes of the OH masers in
the lower-mass stars with a brief evolutionary phase after thermal
pulses of duration $\approx$500 years, when mass-loss rates are
sufficiently high to drive a wind able to host masers. The decline of
the masers would then be a result of rapidly declining mass-loss
rates.

We observed the infrared counterparts of most OH masers in the Arecibo
sample (Jim\'enez-Esteban et al. \cite{Jimenez05}) and monitored them
for several years, including the five stars discussed by Lewis
(\cite{Lewis02}). \object{IRAS~18455+0448} is non-variable,
corroborating Lewis' conclusion that this star is already in the
post-AGB phase. \object{IRAS~20547+0247} (= \object{U~Equ}) is also
non-variable; this is known as a peculiar star surrounded by an
edge-on disk or torus (Barnbaum et al.  \cite{Barnbaum96}), instead of
a radial symmetric CSE. Thus, this star may not be representative of
OH/IR stars in general. \object{IRAS~15060+0947}, as discussed by
Lewis (\cite{Lewis02}), and the remaining two stars,
\object{IRAS~19479+2111} and \object{IRAS~19529+3634},
are large-amplitude variables. We did not find variations in the mean
magnitudes or colors during the monitor period 1999--2005 for any of
these stars, indicating that their mass-loss rates did not change
strongly.  Thus, there is no independent evidence for a change in the
mass-loss process being a cause for the disappearance of the
masers. Furthermore, the typical time a 1~$M_{\sun}$ star spends on
the thermal pulsing AGB is of the order of $5 \cdot 10^5$ years
(Vassiliadis \& Wood \cite{Vassiliadis93}), with at least 10\% of this
time having sufficiently high mass-loss rates to be able to sustain a
maser.  With a OH maser detection rate of 60\% (Lewis \cite{Lewis92})
among IRAS selected AGB star samples, the expected lifetimes of OH
masers are $\ga$30\,000 years, and therefore it is unlikely that the
(temporary) disappearance of the masers in the bluer envelopes is
linked to evolution.  The reappearance of the OH maser in
\object{IRAS~19479+2111} after a couple of years indicates that the
maser extinction in blue envelopes in general is not permanent, but
rather a temporary effect.

The lack of evidence for major changes of the infrared properties
precludes the lack of pumping photons as cause for the (temporary)
extinction of the OH masers in blue envelopes. Variations of density
or velocity coherence disruption remain as alternatives.  One might
envisage modulations of the mass-loss process on timescales of
hundreds of years, which might suffice to affect the excitation
conditions significantly. Such modulations of the envelope structure
were found in studies of dust scattered light at distances up to
$\ga$10$^{21}$ cm from the stars (Mauron \& Huggins \cite{Mauron06}).
At such a range of distances, the history of the mass-loss process can be
studied over $\approx$10\,000 years.  They find in the case of
IRC+10216, a carbon star on the AGB, shells spaced at irregular
intervals corresponding to timescales of 200--800 years (Mauron \&
Huggins \cite{Mauron00}), showing the presence of modulation on the
required timescales. In hydrodynamical models for dust driven AGB
winds, Simis et al. (\cite{Simis01}) found that such quasi-periodicity 
develops naturally in the winds, if dust and gas are not perfectly
coupled. While the increase of densities probably associated with
the shells would improve the excitation conditions for OH masers,
an increase of turbulent motion would likely decrease the gain lengths.

If such modulations of the envelope structure are responsible for OH maser
extinction, then the different lifetimes derived from the Arecibo
sample and from the sample studied here, point to a different
susceptibility of their masers to the inferred changes of the
excitation conditions. The masers in classical OH/IR stars, as studied
in this paper, would be more robust against extinction than the ones
in bluer envelopes. The inability to distinguish OH/IR stars and
`OH/IR star mimics' by ways other than by their masers would then be
explained by the instability of the OH masers, which turn on or off in
response to variations of the envelope structure. The timescales of such
variations would be of the order of $\la 1000$ years in the case of
stars with blue envelopes and $\ga 3000$ years for the reddest OH/IR
stars. Any sample of oxygen-rich AGB stars would then show, at
different times, a different set of stars exhibiting OH maser emission.

\section{Conclusions} 
We find that the lifetime of OH masers in classical OH/IR stars is
$>2800$ years, in contrast to the lifetime implied by the observed
disappearance of several masers in AGB stars with bluer envelopes.
The lower limit of the lifetime is of the order of the wind crossing
time in the CSEs, implying that, in general, no drastic variations in
the structure of the CSEs happen on such timescales, or shorter
ones. If non-variable OH/IR stars are in transition to the post-AGB
stage, a sudden decline of the mass-loss rates at the end of AGB
evolution is ruled out.  The previous observed disappearance of OH
masers in stars with bluer envelopes is probably not associated with
major changes in the mass-loss process. The susceptibility of masers
to smaller variances in their environment may lead to OH masers as
transient phenomena on the AGB.  This gives a natural explanation for
the detection of OH masers in only a part of any AGB star sample with
otherwise similar properties.

\begin{acknowledgements}
We thank B.M. Lewis for information on the most recent observations of
the OH maser emission in several `dead OH/IR stars'.  The comments of
the referee H. Habing are acknowledged.  This research has made use of
the SIMBAD database, operated at CDS, Strasbourg, France. The
observations were made with the Effelsberg 100-m telescope operated by
the Max-Planck-Institut f\"ur Radioastronomie (MPIfR).
\end{acknowledgements}


\longtab{1}{
\begin{longtable}{lccrrrrl}
\caption{\label{sample} OH/IR star sample and 1612 MHz OH maser properties. \\
Coordinates: a) Bowers \& de Jong (\cite{Bowers83a}); b) IRAS;
                      c) Bowers et al. (\cite{Bowers81}); d) this paper; otherwise MSX6C.}\\
\hline\hline\noalign{\smallskip}
Object & \multicolumn{2}{c}{Coordinates (2000)} & v$_b$ & SI$_b$ & v$_r$ & SI$_r$ & Comments \\
       &          &         &  [\kms] & [Jy*\kms] & [\kms] & [Jy*\kms] \\
\noalign{\smallskip}\hline\noalign{\smallskip}
\endfirsthead
\caption{continued.}\\
\hline\hline\noalign{\smallskip}
Object & \multicolumn{2}{c}{Coordinates (2000)} & v$_b$ & SI$_b$ & v$_r$ & SI$_r$ & Comments\\
       &          &         &  [\kms] & [Jy*\kms] & [\kms] & [Jy*\kms] \\
\noalign{\smallskip}\hline\noalign{\smallskip}
\endhead
\hline
\endfoot
OH~10.5+4.5$^a$  & 17 51 48.8  & $-$17 42 30 &  $-$72.2 &    3.3 &  $-$42.5 &    2.8 & \\ 
OH~10.9+1.5      & 18 03 38.9  & $-$18 41 10 &  111.2 &    1.7 &  145.2 &    0.7 & \\ 
OH~11.5+0.1      & 18 10 38.7  & $-$18 52 57 &   10.1 &   23.1 &   75.1 &  130.9 & \\ 
OH~12.3$-$0.2      & 18 13 08.8  & $-$18 27 49 &   21.4 &    9.7 &   51.4 &   11.8 & \\ 
OH~12.8$-$0.9      & 18 16 49.3  & $-$18 15 02 &  $-$68.9 &    4.5 &  $-$43.1 &    5.5 & \\ 
OH~12.8+0.9      & 18 10 06.1  & $-$17 26 34 &   15.1 &    3.6 &   37.1 &    0.8 & \\ 
OH~12.8$-$1.9      & 18 20 37.0  & $-$18 47 11 &  $-$12.8 &   33.4 &   35.1 &   15.6 & \\ 
OH~13.1+5.0      & 17 55 45.1  & $-$15 03 42 &  $-$85.3 &   13.8 &  $-$47.3 &    6.0 & \\ 
OH~13.4+0.8      & 18 12 03.4  & $-$16 55 31 &  $-$22.5 &    1.2 &   15.4 &    3.4 & \\ 
OH~15.4$-$0.1      & 18 18 54.0  & $-$15 34 39 &  $-$44.6 &    3.1 &  $-$12.4 &    4.2 & \\ 
OH~15.4+1.9      & 18 11 34.0  & $-$14 39 54 &   $-$3.4 &    5.3 &   27.2 &    3.6 & \\ 
OH~15.7+0.8      & 18 16 25.7  & $-$14 55 15 &  $-$16.7 &   44.0 &   15.1 &   33.7 & \\ 
OH~16.1$-$0.3      & 18 21 06.9  & $-$15 03 22 &   $-$3.1 &   45.1 &   44.4 &   41.9 & \\ 
OH~16.1+1.4      & 18 14 51.5  & $-$14 16 19 &   28.3 &    1.0 &   66.5 &    0.7 & \\ 
OH~17.0$-$0.1      & 18 21 49.9  & $-$14 10 35 &   32.2 &    6.6 &   68.1 &    4.9 & \\ 
OH~17.1$-$1.2      & 18 26 15.7  & $-$14 42 26 &   $-$9.8 &    3.9 &   17.6 &    4.1 & \\ 
OH~17.2$-$1.1      & 18 26 10.4  & $-$14 28 19 &  157.2 &    1.3 &  184.2 &    1.7 & \\ 
OH~17.4$-$0.3      & 18 23 35.0  & $-$13 55 50 &   10.5 &   12.0 &   47.2 &    6.9 & \\ 
OH~17.7$-$2.0      & 18 30 30.7  & $-$14 28 58 &   46.4 &   52.2 &   75.1 &  131.0 & \\ 
OH~18.2+0.5      & 18 21 56.2  & $-$12 55 22 &  111.9 &    1.8 &  138.4 &    2.4 & \\ 
OH~18.3+0.1$^a$  & 18 23 50.8  & $-$12 50 38 &    9.8 &     -- &    --  &    --  & interstellar\\
OH~18.3+0.4      & 18 22 43.1  & $-$12 47 42 &   31.3 &    8.5 &   64.3 &   10.3 & \\ 
OH~18.5+1.4      & 18 19 35.5  & $-$12 08 09 &  164.6 &   10.4 &  187.8 &    9.4 & \\ 
OH~18.7+1.6      & 18 19 40.6  & $-$11 53 00 &  $-$18.1 &    2.9 &   16.2 &    1.8 & \\ 
OH~18.8+0.4      & 18 24 05.3  & $-$12 26 14 &   $-$3.5 &   38.9 &   28.4 &   22.9 & \\ 
OH~19.2$-$1.0      & 18 29 28.6  & $-$12 37 55 &   29.9 &    8.4 &   68.3 &   12.7 & \\ 
OH~19.5+4.0      & 18 12 22.7  & $-$10 02 50 &   28.5 &    3.3 &   62.6 &    4.6 & \\ 
OH~20.2$-$0.1      & 18 28 13.9  & $-$11 16 10 &    8.5 &   10.2 &   44.3 &    8.8 & \\ 
OH~20.3$-$1.5      & 18 33 34.2  & $-$11 56 45 &  $-$25.1 &    1.7 &    2.2 &    3.1 & \\ 
OH~20.4$-$0.3      & 18 29 35.5  & $-$11 15 54 &      -- &     -- &   60.3 &     -- & \\ 
OH~20.4+1.4      & 18 23 28.7  & $-$10 27 21 &   19.4 &    0.8 &   57.0 &    0.7 & \\ 
OH~20.6+0.3      & 18 27 56.3  & $-$10 46 54 &   69.5 &    2.6 &  112.2 &    2.7 & \\ 
OH~20.7+0.1      & 18 28 30.8  & $-$10 50 52 &  116.9 &   20.3 &  155.7 &   14.2 & \\ 
OH~20.8$-$0.8      & 18 31 23.0  & $-$11 09 54 &   30.3 &    0.9 &   64.9 &    2.6 & \\ 
OH~20.8+3.1      & 18 17 58.5  & $-$09 18 31 &   11.2 &    3.7 &   41.5 &    4.4 & \\ 
OH~21.5+0.5      & 18 28 30.9  & $-$09 58 15 &   95.2 &   24.6 &  135.6 &   18.0 & \\ 
OH~21.9+0.4      & 18 29 30.7  & $-$09 39 50 &   65.8 &    4.8 &  105.9 &    3.6 & \\ 
OH~22.1$-$0.6      & 18 33 34.5  & $-$09 57 36 &  102.4 &    6.9 &  131.1 &    3.1 & \\ 
OH~22.3$-$2.5      & 18 40 38.5  & $-$10 42 35 &   97.5 &    1.2 &  126.9 &    1.4 & \\ 
OH~23.1$-$0.3      & 18 34 11.3  & $-$08 58 02 &   18.7 &   12.8 &   50.6 &   11.7 & \\ 
OH~23.7+1.2      & 18 30 06.9  & $-$07 36 50 &  $-$15.6 &    8.4 &   15.9 &    6.3 & \\ 
OH~23.8+0.2      & 18 33 49.6  & $-$08 04 01 &   93.3 &    3.5 &  118.0 &    9.0 & \\ 
OH~23.8$-$1.1      & 18 38 39.8  & $-$08 41 12 &   32.4 &    2.9 &   66.8 &    7.8 & \\ 
OH~24.3+0.7      & 18 33 09.8  & $-$07 27 58 &   39.3 &    1.8 &   75.8 &    2.4 & \\ 
OH~24.5+0.3      & 18 35 19.0  & $-$07 19 23 &  $-$87.3 &    6.8 &  $-$58.1 &    6.6 & \\ 
OH~24.7$-$0.1      & 18 36 45.9  & $-$07 18 17 &   60.3 &    5.3 &  110.8 &    7.7 & \\ 
OH~24.7+0.1      & 18 35 57.4  & $-$07 18 57 &   39.3 &    5.1 &   73.3 &    4.8 & \\ 
OH~24.7+0.3      & 18 35 29.2  & $-$07 13 11 &   19.8 &   12.8 &   65.0 &   14.8 & \\ 
OH~24.7$-$1.7      & 18 42 22.0  & $-$08 05 00 &   77.0 &    3.1 &  106.3 &    3.1 & \\ 
OH~25.1$-$0.3      & 18 38 15.4  & $-$07 09 54 &  128.7 &    4.8 &  156.2 &    4.1 & \\ 
OH~25.5$-$0.3      & 18 38 50.5  & $-$06 44 50 &    8.4 &    8.3 &   77.0 &   14.7 & \\ 
OH~25.5+0.4      & 18 36 44.0  & $-$06 27 24 &   21.4 &    4.7 &   56.2 &    3.1 & \\ 
OH~26.2$-$0.6      & 18 41 14.4  & $-$06 15 01 &   48.4 &   23.9 &   95.3 &   20.3 & \\ 
OH~26.3+0.1      & 18 38 38.8  & $-$05 49 10 &  $-$41.7 &    1.4 &  $-$11.2 &    1.8 & \\ 
OH~26.4$-$1.9      & 18 46 26.8  & $-$06 40 33 &   13.9 &    9.6 &   44.9 &   17.3 & \\ 
OH~26.4$-$2.8      & 18 49 16.8  & $-$07 13 47 &  $-$79.7 &    3.4 &  $-$53.3 &    3.6 & \\ 
OH~26.5+0.6      & 18 37 32.5  & $-$05 23 59 &   10.5 &  139.4 &   43.5 &  237.9 & \\ 
OH~27.0$-$0.4      & 18 42 01.9  & $-$05 21 07 &   86.0 &    5.2 &  117.7 &    7.0 & \\ 
OH~27.2+0.2      & 18 40 16.4  & $-$05 02 39 &   71.6 &    1.5 &  113.4 &    2.0 & \\ 
OH~27.3+0.2      & 18 40 21.9  & $-$04 57 11 &   36.3 &   38.8 &   64.5 &   19.7 & \\ 
OH~27.5$-$0.9      & 18 44 41.6  & $-$05 09 18 &   92.1 &    7.1 &  121.2 &    9.1 & \\ 
OH~27.8$-$1.5      & 18 47 37.8  & $-$05 11 08 &   67.8 &    2.6 &  102.2 &    2.8 & \\ 
OH~28.5$-$0.0      & 18 43 25.8  & $-$03 55 55 &   92.9 &   13.8 &  122.0 &   13.8 & \\ 
OH~28.7$-$0.6      & 18 45 48.3  & $-$04 00 46 &   27.6 &    8.5 &   65.1 &    4.8 & \\ 
OH~29.4$-$0.8      & 18 47 49.9  & $-$03 29 31 &  106.1 &   17.0 &  140.1 &   16.3 & \\ 
OH~30.1$-$0.2      & 18 47 09.8  & $-$02 35 36 &   31.4 &   16.8 &   69.5 &   16.4 & \\ 
OH~30.1$-$0.7      & 18 48 42.0  & $-$02 50 29 &   76.8 &   65.8 &  121.2 &   79.9 & \\ 
OH~30.7+0.4      & 18 45 52.4  & $-$01 46 43 &   47.8 &   10.0 &   85.2 &    7.6 & \\ 
OH~31.0+0.0      & 18 47 41.2  & $-$01 45 11 &   26.7 &    0.2 &   41.7 &   12.9 & \\ 
OH~31.0$-$0.2      & 18 48 43.1  & $-$01 48 30 &  110.0 &   45.8 &  140.0 &    3.9 & \\ 
OH~31.5$-$0.1      & 18 49 05.0  & $-$01 17 02 &   19.4 &    3.3 &   50.9 &    2.8 & \\ 
OH~31.7$-$0.8      & 18 52 01.4  & $-$01 26 47 &   64.8 &    1.3 &   92.8 &    2.1 & \\ 
OH~32.0$-$0.5      & 18 51 26.2  & $-$01 03 52 &   54.0 &   24.5 &   98.2 &   13.8 & \\ 
OH~32.1+0.9      & 18 46 38.7  & $-$00 17 14 &  123.8 &    1.6 &  150.8 &    2.8 & \\ 
OH~32.8$-$0.3      & 18 52 22.2  & $-$00 14 11 &   42.9 &   55.8 &   78.5 &   26.8 & \\ 
OH~33.4$-$0.0      & 18 52 27.9  & +00 27 30 &   43.2 &    4.3 &   78.3 &    6.2 & \\ 
OH~34.7+0.9      & 18 51 15.9  & +01 56 19 &   10.2 &    4.2 &   44.6 &    5.1 & \\ 
OH~34.9+0.8      & 18 52 15.3  & +02 03 48 &   52.8 &    2.1 &   84.9 &    1.4 & \\ 
OH~35.2$-$2.6      & 19 05 02.0  & +00 48 51 &   21.3 &   21.8 &   73.8 &   15.8 & \\ 
OH~35.6$-$0.3      & 18 57 27.4  & +02 12 18 &   63.0 &   30.2 &   93.0 &   21.1 & \\ 
OH~36.4+0.3      & 18 56 36.4  & +03 06 53 &   82.2 &    3.3 &  122.4 &    2.9 & \\ 
OH~36.9+1.3      & 18 54 16.0  & +04 02 32 &  $-$20.8 &    3.4 &   $-$3.7 &    3.2 & \\ 
OH~37.1$-$0.8      & 19 02 06.3  & +03 20 16 &   73.0 &   10.7 &  103.9 &    9.1 & \\ 
OH~37.7$-$1.4      & 19 05 09.7  & +03 40 58 &   99.2 &    1.3 &  123.9 &    0.6 & \\ 
OH~39.6+0.9A     & 19 00 55.5  & +06 21 05 &  \multicolumn{4}{c}{no detection} &IRAS~18584+0616 \\
OH~39.6+0.9B$^b$ & 19 00 24.6  & +06 16 31 &  \multicolumn{4}{c}{no detection} &IRAS~18579+0612 \\ 
OH~39.7+1.5      & 18 58 30.1  & +06 42 55 &    2.1 &   75.6 &   38.0 &  129.7 & \\ 
OH~39.9$-$0.0      & 19 04 09.6  & +06 13 16 &  132.9 &   10.6 &  165.0 &   11.3 & \\ 
OH~40.1+2.4      & 18 55 56.9  & +07 30 28 &   26.0 &    1.1 &   67.8 &    1.5 & \\ 
OH~42.3$-$0.1      & 19 09 08.2  & +08 16 34 &   41.4 &   22.7 &   77.6 &   41.8 & \\ 
OH~42.6+0.0      & 19 08 58.4  & +08 37 49 &   34.1 &    5.6 &   71.8 &    4.8 & \\ 
OH~42.8$-$1.0$^d$  & 19 13 36.4  & +08 22 39 &  $-$29.5 &    2.2 &   $-$1.8 &    1.0 & Pos. error: $\approx$2\arcmin\\ 
OH~43.6$-$0.5      & 19 12 46.9  & +09 18 24 &   58.8 &    1.6 &   86.3 &    1.4 & \\ 
OH~43.8+0.5      & 19 09 31.8  & +09 51 50 &   $-$4.3 &    5.8 &   22.5 &    4.4 & \\ 
OH~43.9$-$1.0      & 19 15 16.4  & +09 15 45 &   37.8 &    0.8 &   65.5 &    1.1 & \\ 
OH~43.9+1.2      & 19 06 42.8  & +10 14 33 &   32.4 &    2.0 &   67.4 &    1.5 & \\ 
OH~44.8$-$2.3      & 19 21 36.6  & +09 27 57 &  $-$88.9 &   10.1 &  $-$53.8 &   14.4 & \\ 
OH~45.5+0.1      & 19 14 19.6  & +11 10 35 &   16.5 &   14.9 &   53.7 &   14.2 & \\ 
OH~51.8$-$0.2      & 19 27 42.1  & +16 37 25 &  $-$18.5 &    7.3 &   23.1 &    5.4 & \\ 
OH~52.4+1.8      & 19 21 31.5  & +18 10 09 &   $-$0.3 &    2.7 &   31.3 &    4.5 & \\ 
OH~53.6$-$0.2      & 19 31 25.3  & +18 13 10 &   $-$4.0 &    9.8 &   25.5 &   23.9 & \\ 
OH~55.0+0.7      & 19 30 29.4  & +19 50 41 &   12.5 &   14.8 &   44.2 &    9.3 & \\ 
OH~57.5+1.8A     & 19 31 38.9  & +22 35 17 &  $-$88.0 &    1.9 &  $-$59.2 &    1.6 & \\ 
OH~57.5+1.8B     & 19 31 45.5  & +22 33 42 &   30.1 &    2.2 &   46.3 &    4.8 & \\ 
OH~63.9$-$0.2      & 19 52 57.9  & +27 07 45 &  $-$11.3 &    1.7 &   22.3 &    1.8 & \\ 
OH~65.5+1.3      & 19 51 21.2  & +29 13 01 &  $-$41.1 &    6.9 &   $-$2.7 &    6.6 & \\ 
OH~65.7$-$0.8      & 19 59 39.4  & +28 23 10 &  $-$70.2 &    2.4 &  $-$46.1 &    4.0 & \\ 
OH~66.8$-$1.3      & 20 04 20.8  & +29 04 07 &  $-$79.1 &    2.6 &  $-$50.2 &    2.7 & \\ 
OH~75.3$-$1.8      & 20 29 08.5  & +35 45 42 &  $-$16.4 &    3.7 &   10.0 &    3.7 & \\ 
OH~77.9+0.2      & 20 28 30.7  & +39 07 00 &  $-$50.9 &    8.5 &  $-$26.0 &    5.6 & \\ 
OH~80.8$-$1.9      & 20 46 25.5  & +40 06 58 &  $-$28.7 & 1151.4 &   25.6 &  215.5 & \\ 
OH~83.4$-$0.9      & 20 50 58.6  & +42 48 12 &  $-$58.1 &   11.6 &  $-$18.1 &    5.2 & \\ 
OH~85.4+0.1      & 20 53 38.0  & +44 58 07 &  $-$36.8 &    3.0 &   $-$7.9 &    2.8 & \\ 
OH~104.9+2.4     & 22 19 27.2  & +59 51 20 &  $-$41.3 &   36.3 &   $-$8.5 &   26.5 & \\ 
OH~127.8+0.0     & 01 33 51.2  & +62 26 53 &  $-$66.5 &   67.3 &  $-$41.8 &   40.9 & \\ 
OH~138.0+7.2$^c$ & 03 25 08.4  & +65 32 07 &  $-$47.7 &   18.5 &  $-$26.5 &    8.9 & \\ 
OH~141.7+3.5$^c$ & 03 33 30.5  & +60 20 09 &  $-$70.2 &    4.9 &  $-$43.7 &    1.7 & \\ 
\end{longtable}
}

\end{document}